\input{aipcheck}

\documentclass[
    ,final            
  ]
  {aipproc}
\layoutstyle{6x9}

\def\MeV{{\rm MeV}}
\def\keV{{\rm keV}}
\def\nostrocostrutto#1\over#2{\mathrel{\mathop{\kern 0pt \rlap
  {\raise.2ex\hbox{$#1$}}}
  \lower.9ex\hbox{\kern-.190em $#2$}}}
\def\gsim{\nostrocostrutto > \over \sim}   

\begin{document}

\title{Searching for the Scalar Glueball}


\author{Wolfgang Ochs}{
  address={Max-Planck-Institut f\"ur Physik, Werner-Heisenberg Institut\\
D-80805 Munich, F\"ohringer Ring 6, Germany}
}

\begin{abstract}
Existence of gluonic resonances is among the early expectations of QCD. 
Today, QCD calculations predict the lightest glueball to be a scalar state
with mass within a range of about 900-1700 MeV but there is no consensus about its
experimental evidence. In a re-analysis of the phase shifts for 
$\pi\pi$ scattering up to 1800 MeV where such states should show up we find  
the broad resonance $f_0(600)/\sigma$ contributing to the full mass range 
and the narrow $f_0(980)$ and $f_0(1500)$ but no evidence for $f_0(1370)$.
Phenomenological 
arguments for the broad state to be a glueball are recalled. It is argued
that the large
radiative width of $f_0(600)/\sigma$ reported recently is not in  contradiction
to this hypothesis but is mainly due to $\pi\pi$-rescattering. The small
``direct'' radiative component is consistent with QCD sum rule predictions for the light
glueball. 
\end{abstract}

\maketitle


\section{Expectations for the lightest glueball in QCD}
Quantitative results on glueballs are available today from \\
1. Lattice QCD: 
The existence of glueballs within the purely 
gluonic theory is established and the
lightest state is found in the scalar sector with mass around 1.7 GeV. 
In full QCD both glue and $q\bar q$ states couple to the flavour singlet
$0^{++}$ states and first ``unquenched'' results for the lightest gluonic
state point towards a lower mass of around 1 GeV \cite{michael}
(recent review \cite{neile}). 
Further studies concerning the dependence on lattice spacing
and the quark mass appear important.\\
2. QCD sum rules: Results on the scalar glueball and various decays are
obtained in \cite{veneziano}. The lightest gluonic state is found in
the mass range (950-1100) MeV with a decay width of $\sim$1000 MeV 
into $\pi\pi$ and the width into $\gamma\gamma$ of (0.2-0.6) keV.  
Other analyses find similar or slightly higher masses (around 1250 MeV)
for the lightest glueball \cite{steele}.
\section{The scalar meson spectrum}
In the search for glueballs one attempts to group the experimentally
observed scalar mesons
into flavour multiplets (either $q\bar q$
or tetraquarks) and to identify supernumerous states.
The existence of such states could be a hint for glueballs either pure or   
mixed with $q\bar q$ isoscalars.
A signature of gluonic states is its abundance 
in so-called ``gluon rich'' processes and their suppression in $\gamma\gamma$
reactions.

In a popular scheme the light scalars $\sigma(600),\ \kappa(800),\
f_0(980),\ a_0(980)$ are put together into one multiplet, either as $q\bar
q$  or $4q$ nonet (see, for example Refs. \cite{scadron},\cite{jaffe}). Then, 
a $q\bar q$ 
multiplet can be formed with the heavier $a_0(1450),~K^*_0(1430)$; with
nearby masses three isoscalars can be found at 
1370, 1500 and 1710 MeV and this suggests their interpretation
as mixtures of the two $q\bar q$ nonet members and one
glueball (for an early reference, see \cite{ac}). 
 
A potential problem in this scheme for the glueball is the very existence of
$f_0(1370)$, otherwise there is no supernumerous state in this mass range.
Some problems with this state will be discussed below, see also the
review \cite{klemptrev}. 
The low mass multiplet depends on the existence
of $\kappa$ which we consider as not beyond any doubt:
the observed related phase motion in $K\pi$ scattering \cite{lass} 
is less pronounced as the corresponding one of $``\sigma"$ in $\pi\pi$
scattering, see Fig. 1, discussed below.

In the scheme we prefer \cite{mo}
the lightest $q\bar q$ nonet contains $f_0(980),~f_0(1500)$ together with
$a_0(1450),~K^*_0(1430)$. The supernumerous state $\sigma/f_0(600)$, called
previously $f_0(400-1200)$, corresponds to a very broad
object which extends from low energy up to about 2 GeV
and is interpreted as largely gluonic. No separate
$f_0(1370)$ is introduced, nor $\kappa(800)$. 
Our classification is consistent with various findings on production and
decay processes including $D,D_s,B$ and $J/\psi$
decays \cite{mo,momontp,mobdecay}. 

Related
schemes are the Bonn model \cite{klempt} with a similar mixing scheme for
the isoscalars and the K-matrix model \cite{anis} which finds a similar  
classification (but with $f_0(1370)$ included) and a broad glueball,    
but centered at the higher masses around 1500 MeV.

\section{Phase shift results on $\pi\pi\to \pi\pi$ up to 1800 MeV}
Our focus here is on the significance of $f_0(1370)$ and the appearance of
$\sigma/f_0(600)$ which was $f_0(400-1200)$ before 
and is sometimes treated
as ``background'' where we discribe results from an ongoing analysis (see also
\cite{womont}).

Information on $\pi\pi$ scattering can be obtained from production
experiments like $\pi p \to \pi \pi n$ by isolating the contribution of the
one-pion-exchange process. In an unpolarised target experiment
these amplitudes can be extracted by using dynamical assumptions, such as
``spin and phase coherence'', which have been tested by experiments with 
polarised target. At the level of the process $\pi\pi\to\pi\pi$ in different
charge states one measures the distribution in scattering angle,
$z=\cos\theta^*$, or their moments $\langle Y^L_M \rangle$, in a sequence of
mass intervals. The $\pi\pi$ 
partial wave amplitudes $S,P,D,F,\ldots$ can be obtained
in each bin from the measured moments up to the overall phase and a discrete
ambiguity (characterised by the ``Barrelet Zeros''). The overall phase can  
be fixed by fitting a Breit Wigner amplitude for the leading resonances     
$\rho,~f_2(1270)$ and $\rho_3(1690)$ to the respective experimental 
moments.

Energy-independent 
phase shift analyses of this type for $\pi^+\pi^-$ scattering have been
performed
by the CERN-Munich group: an analysis guided by a global resonance fit (CM-I   
\cite{cm}) and an analysis to reveal all ambiguities 
by CM-II \cite{cm2} and by   
Estabrooks and Martin \cite{em}; 
4 different solutions have been found above 1 GeV in mass. Up to 1400 MeV a
unique solution has been established \cite{kpy} using results from polarised
target 
and unitarity. Two solutions remain above 1400 MeV,
classified according to Barrelet zeros in \cite{cm2} as ($---$) and ($-+-$)
corresponding to sols. A,C in \cite{em}.

\begin{figure*}[t]
\begin{tabular}{@{}lll}
\includegraphics*[width=5.0cm,bbllx=3.0cm,bblly=1.8cm,bburx=9.9cm,
bbury=8.2cm]{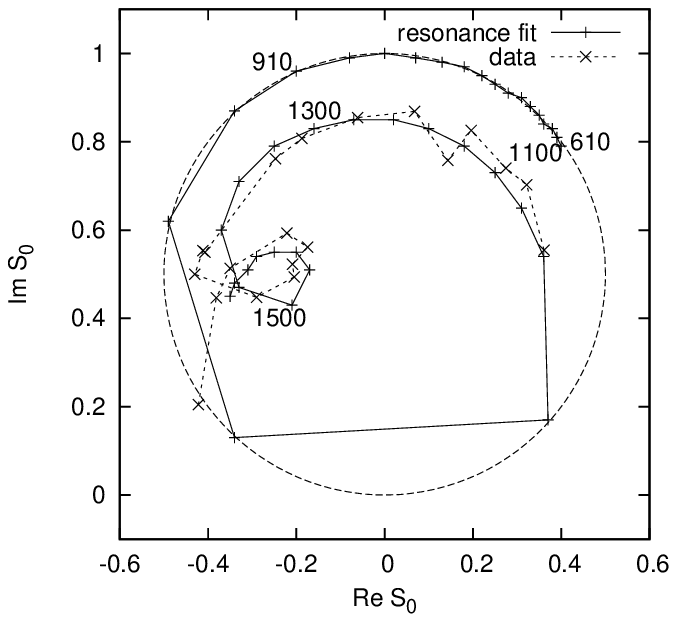} \quad 
\includegraphics*[width=5.0cm,bbllx=3.0cm,bblly=1.8cm,bburx=9.9cm,
bbury=8.2cm]{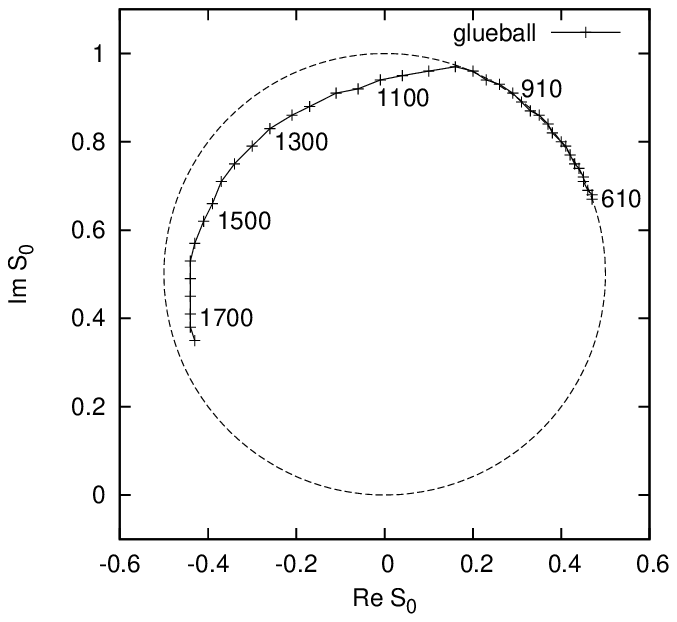}\\
\end{tabular}
\caption{Argand diagram for the corrected partial wave $S_0$ 
(CM-I/II) in comparison with the resonance fit in Eq. (1); right panel: 
broad component $f_0(600)/\sigma$ from the fit.}
\label{fig:resonances}
\end{figure*}

The remaining ambiguity has been resolved recently in \cite{womont}
by comparison with the isoscalar S wave $S_0$ reconstructed from the
$\pi^+\pi^-\to
\pi^0\pi^0$ data (GAMS collaboration \cite{gams}) 
and results on $I=2$ scattering.
The $S_0$ wave obtained shows a
qualitatively similar behaviour to the $S_0$ solution 
above. In particular, 
both solutions show an $f_0(1500)$ resonance circle
in the complex plane (Argand diagram)  above a slowly
moving circular background amplitude. The $S_0$($-+-$) amplitude is shown in 
Fig. \ref{fig:resonances}.

The resulting amplitude $S_0(-+-)$ 
is shown in Fig. \ref{fig:resonances} using the CM-II data after    
correction for the more recent $I=2$ amplitudes. 
The curves refer to a fit of the data (CM-II for $M_{\pi\pi}>1$ GeV,
CM-I for $M_{\pi\pi}<1$ GeV)
to a unitary S-matrix in the space of 3 reaction channels
($\pi\pi,K\bar K, 4\pi$) as product of individual S-matrices for resonances
$S_R=1+2i T_R$
\begin{eqnarray}
S&=&S_{f_0(980)}S_{f_0(1500)}S_{\rm broad}\\
T_R&=& \rho^{\frac{1}{2}T} (g_ig_j)\rho^\frac{1}{2}~~ 
     [M_0^2-M_{\pi\pi}^2-i(\rho_1 g_1^2 +\rho_2 g_2^2+ \rho_3 g_3^2)]^{-1}
    \label{smatrix}
\end{eqnarray}
where $\rho_i=2k_i/\sqrt{s}$. This simplified amplitude 
for the unitary S matrix is a generalisation of the so-called
Dalitz-Tuan form: it is well suited to locally describe the superposition
of a smooth background with a narrow resonance.
The fit including 3 resonances gives a reasonable description of the 
measured inelasticities $\eta$ and phase shifts $\delta$.
For $f_0(1500)$ the fit parameters $M_0= 1510~{\rm MeV},~\Gamma_{\rm tot}=
88~{\rm MeV},~B(f_0\to\pi\pi)=38\%$ are obtained,  comparable to the 
PDG results. This is remarkable as the latter values come from the
$\pi\pi$
fraction of all observed decay rates but our value from elastic scattering. 

The broad object is also described
by a Breit-Wigner form (\ref{smatrix}) with parameters \cite{womont}
\begin{equation}
M_{\rm BW} \sim 1100~\MeV,\quad \Gamma_{\rm BW} \sim 1450~ \MeV.
\label{bwmass}
\end{equation}
 The elastic width is about  85\% whereas the GAMS data
suggest rather a smaller value around 70\%. This should be considered as
systematic uncertainty. More details will be given
elsewhere. This amplitude is shown in Fig.
\ref{fig:resonances}, right panel. It completes about 3/4 of the full resonance circle.
The parameter $M_{\rm BW}$ 
refers to the mass where the amplitude is purely
imaginary. It may be  different from the pole mass which is referred to as
resonance mass. The determination of this mass requires 
an analytic continuation of the propagator into the deep complex region 
and in many analyses it appears to be considerably smaller. 

The data in Fig. \ref{fig:resonances} suggest the existence of a 
broad state in $\pi\pi$ scattering,
centered around 1000 MeV along the physical region 
and what is called $f_0(600)$ or $\sigma$ refers to the same state.
A large difference between both masses is revealed
in a simple analytical model (one channel, one pole) \cite{mno} in extension
of the model \cite{mennessier} where the propagator real part is calculated
from a dispersion relation. The model is fit to the $\pi\pi$ data at lower
energies ($<700$ MeV), then 
 the ``on-shell'' mass $M^{\rm os}$ 
where the amplitude is purely imaginary (phase shift
at 90$^\circ$), the pole mass $M^{pole}$ 
and the corresponding widths are obtained as
\begin{equation}
 M^{\rm os}=920~\MeV,~  \Gamma^{\rm os}=1020~ \MeV; \qquad 
M^{\rm pole}=422~\MeV, ~ \Gamma^{\rm pole}=580~ \MeV.  \label{sigmas}
\end{equation}   
The on shell results from this low energy fit resemble 
the Breit Wigner parameters
in (\ref{bwmass}). 
The very different pole and on-shell masses are related by the
analytic extrapolation and refer to the same state. On the other hand, fits
with a pole mass near 1000 MeV are possible as well as shown at this
conference \cite{mink}.

We note that the data presented in Fig. \ref{fig:resonances} 
do not give any indication of 
the existence of $f_0(1370)$ which is expected to show up as
a second circle in the Argand diagram besides $f_0(1500)$ 
with respective signals 
in $\eta^0_0$ and $\delta_0^0$. In fact, 
none of the energy-independent bin by bin phase shift analyses of the CM
or CKM data \cite{cm,em,cm2,kpy} nor of the GAMS data \cite{gams,womont} 
gave such an indication. From our analysis 
we exclude an additional state  with branching ratio
$B(f_0(1370)\to \pi\pi) 
> 0.1$ near 1370 MeV (this would correspond to
a circle of  diameter 0.1). It should be noted that the ``experimental'' 
bin-by-bin $\eta^0_0$ and $\delta_0^0$ values
are obtained from the original moments in very good
fits ($\chi^2$/data point = $0.\sim 1.$). 

The existence of $f_0(1370)$ is still 
controversially discussed and very different
views and numbers are recommended. At this conference 
two other analyses \cite{bugg,oller} show this. In the analysis \cite{bugg} 
CM-I moments 
have been fitted directly by a model
amplitude with resonances in all relevant partial waves ($\chi^2$/data
point $\gsim$ 2).
The resulting amplitude $S_0$ includes $f_0(1370)$ which appears
with  extra circle of diameter 0.25 in the Argand diagram near the mass of 1300
MeV. Such an effect is hard to reconcile with any of the energy independent phase
shift analyses. Another presentation \cite{oller} 
based on phase shift results does not produce any extra circle in the 1300
MeV region. Above 1400
MeV a set of phase shifts from CKM is used, different from 
ours, which does not 
reveal any extra circle at all, neither for $f_0(1500)$ nor for $f_0(1370)$ in the mass
range below 1800 MeV. 
The $f_0(1370)$ phenomenon is clearly of a different nature as in Ref. \cite{bugg}. 
The ambiguity in the phase shifts at the higher energies in both analyses
\cite{womont} and \cite{oller} needs further study,
in particular, the consistency with the GAMS data \cite{gams}.

\section{Glueball interpretation of the broad object $f_0(600)$}
{\it Arguments in favour of glueball} have been put forward in 
\cite{mo,momontp,mobdecay}.\\
1. This state is produced in almost all ``gluon rich'' processes, including
central production $pp\to p(\pi\pi)p$, $p\bar p\to 3\pi$, $J/\psi\to
\gamma \pi\pi(?)$, $\gamma K\bar K,\gamma 4\pi$, $\psi'\to \psi\pi\pi$,
$\Upsilon'',\Upsilon'\to \Upsilon \pi\pi$ and finally $B\to K\pi\pi,
B\to K\bar K K$ related to $b\to sg$. The high mass tail above 1 GeV is seen
as ``background'' in $J/\psi \to \gamma K\bar K$ and in $B$ decay
channels where it leads to striking interference phenomena with
$f_0(1500)$ \cite{mobdecay}. Only the channel  $J/\psi\to \gamma \pi\pi$
is problematic.\\
2. Within our classification scheme \cite{mo} 
without $f_0(1370)$ the state
$f_0(600)$ is supernumerous.\\
3. The mass and large width is in agreement with the QCD sum rule results
(see below) and the mass also with the first results 
from unquenched lattice QCD.\\
4. Suppression in $\gamma\gamma$ production.\\
Recently, the radiative width $\Gamma(f_0(600)\to\gamma\gamma)=(4.1\pm 0.3)$
keV has been determined by Pennington \cite{pennington} from the process
$\gamma\gamma\to \pi\pi$. As this
number is larger than expected for glueballs,  he concluded 
this state ``unlikely to be gluonic''. 
A resolution of this conflict has been suggested recently \cite{mno} as
follows.

{\it The physical processes in 
$\gamma\gamma\to \pi\pi$}  
at low energies are different from the ones at high energies. At low energies,
the photons couple to the charged pions 
and the Born term with one pion exchange dominates
in $\gamma\gamma\to \pi^+\pi^-$, in addition there is a contribution from 
$\pi^+\pi^-$ rescattering. Explicit models with $\pi\pi$ scattering as input
and with $\sigma/f_0(600)$ pole, can explain the low energy
processes
\cite{mennessier}, also calculations in $\chi PT$ with
non-resonant $\pi\pi$ scattering at low energies.
In this case of the rescattering contribution, a resonance
decaying into $\pi\pi$ would also decay into $\gamma\gamma$ irrespective of
the constituent nature of the state.

At high energies, the photons do resolve the constituents of the produced
resonances: for example, the radiative widths of tensor mesons $f_2,f_2',a_2$
in the region 1200-1500 MeV follow the expectations
from a $q\bar q$ state; the rescattering contibution for $f_2\to
\gamma\gamma$ is limited to be lower than 10-20\% \cite{mennessier}.
   
The model by Mennessier \cite{mennessier} satisfies the constraints 
from unitarity (Watson theorem and generalisations) and analyticity. 
The dispersion relations create a
polynomial ambiguity and this allows introducing an arbitrary 
``direct'' coupling of
resonances into two photons besides the coupling of photons to charged pions
and the coupling of hadrons among themselves within a field theoretic
approach. The case of tensor mesons (and others) suggest 
attributing the direct terms to parton annihilation processes. 

In our application to the $\sigma/f_0(600)$ resonance we restrict the
analysis to the $\pi\pi$ channel only in a mass range $m_{\pi\pi}<700$ MeV. 
Once the parameters of $\pi\pi\to\pi\pi$ scattering are determined the
 $\gamma\gamma \to \pi\pi$ processes in both charge states are calculable 
as superposition of Born term, rescattering
and direct contribution with the direct coupling as only free parameter.
As a result we obtain \cite{mno}
\begin{equation}
\Gamma_{\sigma\to\gamma\gamma}^{\rm dir}\simeq (0.13\pm 0.05)~{\rm
keV}~,\quad
\Gamma_{\sigma\to\gamma\gamma}^{\rm resc}\simeq (2.7\pm 0.4)~{\rm keV}~;
\label{raddir}
\end{equation}
this corresponds to the total radiative width of 
$ \Gamma_{\sigma\to\gamma\gamma}^{\rm tot}\simeq (3.9\pm 0.6)~{\rm keV}$
which is compatible with the range 1.2 $\sim$ 4.1 keV obtained in other
analyses.
The direct radiative width in Eq. (\ref{raddir}) is then to be compared with
the predictions for different intrinsic structures of this resonance. 

In comparing with the predictions from the QCD sum rules it is more
appropriate to consider on shell ``physical'' quantities determined
from the physical region along the real axis. As the resonance mass is
above the mass region fitted these numbers should be considered as crude
approximation. The predictions for the lightest
gluonium state are quoted in \cite{mno} as
\begin{equation}
{\rm gb:} ~~M \simeq (950 \sim 1100)~ \MeV,~~ \Gamma \simeq 1050~ \MeV,~~ 
\Gamma_{\gamma\gamma}\simeq (0.2\sim 0.6)~ \keV.
\label{qssr}
\end{equation}
There is a remarkable agreement with the results from our
analysis for
the on shell quantities in (\ref{sigmas}) and with the direct decay width
$\Gamma^{\rm os,dir}_{\gamma\gamma}=(1.0\pm0.4)$ keV (see also Eq.
(\ref{raddir})). 
On the other hand, a large $q\bar q$ or $4q$ component is found disfavoured.

In conclusion, the broad state $f_0(600)/\sigma$ 
observed in $\pi\pi$ scattering with maximal amplitude around 1 GeV
 and in other
channels is a good glueball candidate. The large width into two photons
is not in contradiction with this view if the large contribution
from $\pi\pi$ rescattering is taken into account. The observed parameters
are in remarkable agreement with QCD sum rule expectations.  


\begin{theacknowledgments}
I would like to thank Peter Minkowski, Gerard Mennessier and Stephan Narison
for the collaboration on the topics of this talk and many discussions.
It is also a pleasure to thank George Rupp and his crew for the organisation
of this enjoyable conference.  
\end{theacknowledgments}



\bibliographystyle{aipproc}   

\bibliography{sample}




\end{document}